\documentclass[12pt]{article}
\begin{document}

\begin{titlepage}
\begin{flushright}
{\large \bf UCL-IPT-97-04,\\ CINVESTAV-FIS-97-03}
\end{flushright}
\vskip 0.5cm
\begin{center}

{\Large \bf Quark mixings as a test of a new symmetry of quark Yukawa 
couplings}
 \vskip 0.6cm

{\large H.~Gonz\'alez$^{1,3}$, 
S.R.\ Ju\'arez W.$^{2}$, P.\ Kielanowski$^{1,4}$\\[4pt]
and  G.~L\'opez Castro$^{5,1}$}\\[5pt]

 {\em $^1$ Departamento de F\'\i sica, CINVESTAV del IPN, Apartado}
\\ {\em Postal 14-740, 070000 M\'exico, D.F. M\'exico} 

{\em $^2$ Escuela Superior de F\'\i s. y Mat. del IPN, U.P.-A.L.M. } 
\\ {\em Edif. 9, 07738 M\'exico, D.F., M\'exico}\\

{\em $^3$ Programa de Mat. y F\'\i s., Universidad Surcolombiana}\\
{\em A.A. 385 Neiva, Colombia}
 
{\em $^4$ Department of Physics, Warsaw University, Poland}

{\em $^5$ Institut de Physique Th\'eorique, Universit\'e catholique de}\\
{\em Louvain, B-1348 Louvain-la-Neuve, Belgium}
\end{center}

\vskip 1.6cm

\begin{abstract}
 Based on the hierarchy exhibited by quarks masses at low energies, we 
assume that Yukawa couplings of up and down quarks are related by $Y_u 
\propto Y_d^2$ at grand unification scales. This ansatz gives rise to 
a symmetrical CKM matrix at the grand unification (GU) scale. 
Using three specific models as illustrative 
examples for the evolution down to low energies, we obtain the entries and 
{\em asymmetries} of the CKM matrix which are in very good agreement with their 
measured values. This indicates that the small asymmetry of the CKM matrix
at low energies may be the effect of the renormalization group evolution only.
\end{abstract}


\end{titlepage}%

\medskip

\newpage

\section{Introduction}
In the Standard Model (SM) of quark and lepton interactions, all 
the fermions get their masses from Yukawa couplings after spontaneous 
breaking of the $SU(2)_L \times U(1)_Y$ gauge symmetry (SSB). 
Since the intensity of Yukawa couplings is described by arbitrary 
complex constants, the quark mass 
matrices induced by SSB are, in general, non-diagonal. The 
diagonalization of these mass matrices yields non-diagonal (in flavor 
space) charged weak interactions, which are described by the unitary 
Cabibbo-Kobayashi-Maskawa (CKM) mixing matrix
 \begin{equation}
V_{ckm}=V_uV_d^{\dagger },
\end{equation}
where $V_u$ ($V_d$) is a unitary matrix that diagonalizes the up (down) 
quark mass matrix.

  Thus, the original information contained in the gauge-invariant Yukawa 
couplings for quarks is drifted to 10 free parameters contained in quark 
masses and mixings at low energies. Therefore, if additional symmetries 
are present in the Yukawa couplings before symmetry breaking, one can 
expect to observe their traces in the structure of quarks masses and 
mixings at low energies. However, 
the values of the Yukawa couplings cannot be
uniquely recovered from the observable quark masses and the parameters of
the CKM matrix. Nevertheless, their measured values can be used as 
useful constrains on the
search for symmetries of Yukawa couplings. In fact, the main motivation of 
the classical Fritzsch's Ansatz \cite{Fritzsch} for
the Yukawa couplings was the phenomenological compatibility with 
the observable values of quark masses and mixings. Fritzsch's Ansatz 
was eventually
excluded by the high mass of the top quark and now the structure of 
the Yukawa couplings for the up and down quarks is expected to be different 
and mainly motivated by grand unified theories (GUT).
 
  Present information about quark masses and mixings \cite{PDG} indicates 
that, at 
low energies, the unitary CKM matrix is almost symmetric. At the same 
time, the diagonal form of up and  down quark mass matrices satisfies the 
approximate relation $M_u \approx M_d^2 m_t/m_b^2$. One may wonder if 
these properties of low energy observables can be correctly reproduced by 
assuming that the CKM matrix is symmetrical and that the relation
\begin{equation}
Y_u\propto Y_d^2
\end{equation}
($Y_{u,d}$ are Yukawa couplings of up and down quarks) are exact 
properties valid at the energy scales of grand unification. In this letter we 
adopt this hypothesis. By assuming the validity of Eq.~(2) at GUT scales, 
we find a symmetrical CKM matrix. The evolution of Yukawa couplings down 
to observable low energies, generates the correct asymmetries and 
absolute values for the entries of the CKM matrix.

To our knowledge, this approach is different from previous studies 
(see for example Refs.~[3--7]).
Current approaches to the problem of the 
generation of quark masses and mixings, are based on the postulation of 
specific Yukawa couplings valid at GUT scales which can be chosen in such 
a way
that the  corresponding low energy data are reproduced. Usually, one can 
further assume some textures (zeros) for entries in the mass matrices in 
order to account for additional symmetries present at GUT scales.

 A different view based on the notion of {\em natural} mass matrices, 
was introduced recently by Peccei and Wang \cite{Pec-W}, \cite{Wang}.
 Their main idea is to derive the GU textures of quark Yukawa couplings 
by evolving to high energies the observed values of quark masses. 
Their naturalness condition is based on the requirement that none of the 
small observables at low energies are derived
by the approximate cancellation of large quantities. As a result they have
obtained several possible scenarios for the GU scale Yukawa couplings.

\section{Quark masses and mixings at 1 GeV}
As already mentioned, the hierarchy observed in low energy values of 
quark masses and mixings can be used as a guide to search for the 
structure of quark Yukawa couplings. This hierarchy can be better 
appreciated by using the small parameter $\lambda \approx 0.22$. The 
measured quark masses below the top mass scale can be written as:

\begin{eqnarray}
M_u &=& m_t \cdot \mbox{Diag}\left( 
\alpha _1^u\lambda ^8,\alpha _2^u\lambda ^4, 1\right) 
\nonumber \\ M_d &=& m_b \cdot \mbox{Diag}\left( 
\alpha _1^d\lambda ^4,\alpha _2^d\lambda ^2, 1\right) 
\end{eqnarray}
where the coefficients $\alpha _i^u$ and $\alpha _i^d$ are of the order 1 
(see for example \cite{Pec-W}).

On the other hand, the hierarchical structure of the CKM matrix is better 
seen in Wolfenstein's parameterization Ref.~\cite{Wolfen} (which is 
unitary up to order $\lambda^4$): \begin{eqnarray}
\left[ V_{ckm}\right]_W=\left( 
\begin{array}{ccc}
1-\frac{1}{2}\lambda^2 & \lambda & \lambda ^3A(\rho -i\eta ) \\ 
-\lambda & 1-\frac 12\lambda ^2 & \lambda ^2A \\ 
\lambda ^3A(1-\rho -i\eta ) & -\lambda ^2A & 1
\end{array}
\right).
\end{eqnarray}
It is worth noting the following properties from Eqs. (3) and (4):
\begin{enumerate}
\item  The up and down quark mass matrices $M_{u,d}$ exhibit the scaling
\begin{equation}
M_u\propto M_d^2.
\end{equation}
\item  With high precision the CKM matrix is almost symmetrical.
In fact 
\begin{equation}
\left| V_{12}\right| ^2-\left| V_{21}\right|^2=
\left| V_{23}\right| ^2-\left| V_{32}\right|^2=
\left| V_{31}\right| ^2-\left| V_{13}\right|^2\sim \lambda ^6.
\end{equation}
where the last equality follows from Eq.~(4) and the other two from unitarity 
of the CKM matrix.
\end{enumerate}

The properties given in Eqs.~(5) and (6) will be the basis of our
further discussion. In our formalism we will generate the symmetrical CKM
matrix at the scale of GU and we will show that upon the evolution with the
method of the Renormalization Group (RG) there will 
appear terms that break the
symmetry of the CKM matrix to the correct order.
\section{
Eigenvalue and eigenvector parameterization of the CKM matrix}
Our formalism of the generation of the CKM matrix will be guided by the
eigenvalue and eigenvector (EE) parameterization of the CKM matrix Ref.~\cite
{Piotr}. In this parameterization the CKM matrix is written in the following
form: 
\begin{equation}
V_{ckm}=\hat AD\hat A^{\dagger }.
\end{equation}
Here $D$ is the diagonal matrix 
\begin{equation}
D=\mbox{Diag}\left( e^{-2\pi i/3},\quad e^{2\pi i/3},\quad 1\right)
\end{equation}
and the matrix $\hat A$ is unitary. Because of the rephasing freedom of
the quark fields, $\hat A$ has only 4 parameters as the original CKM\ 
matrix.
In practice, it turns out that a two-angle parameterization of the real 
matrix $\hat A$ , namely 
\begin{equation}
\hat A=\left( 
\begin{array}{ccc}
c_1&-s_1&0 \\ 
s_1c_2&c_1c_2&-s_2 \\ 
s_1s_2&c_1s_2&c_2
\end{array}
\right)
\end{equation}
where $c_i=\cos\beta_{i}$, $s_i=\sin\beta_{i}$, is good enough to have a 
reasonable agreement with data (the fit to entries of the CKM matrix 
gives: $\beta _1=0.1293\pm 0.0010,\quad \beta _2=0.0239\pm 0.0017,\quad \chi
^2=8.73$. See also Ref. \cite{xing}).

 The matrix $\hat A$ in Eq.~(7) can be interpreted as a universal
matrix that diagonalizes the mass matrices of up and down quarks. For this
to be true a rephasing of quark fields according to Eq.~(12) (below) is 
required before diagonalization of mass matrices (otherwise the CKM 
matrix would become the unit matrix).
After this rephasing of quark fields, the mass matrices in weak 
($\widetilde{M_i}$) and mass eigenstates ($ M_i$) are related by 
\begin{equation}
M_u=\hat A\widetilde{M_u}\hat A^{\dagger },\quad M_d=\hat A\widetilde{M_d}%
\hat A^{\dagger }.
\end{equation}
 Since $\widetilde{M_i}$ are proportional to Yukawa coupling matrices 
$Y_i$, Eqs. (10) and (5) would lead to the relation 
\begin{equation}
Y_u=C\ Y_d^2\label{eq11}
\end{equation}
where $C$ is some proportionality constant (see Eq.~(20) below). 
This means 
that the Yukawa couplings 
of up quarks are simple functions of the Yukawa couplings of down quarks.

To conclude this section let us observe that in our scenario the 
$\hat A$ matrix given in Eq.~(9) is real. The corresponding CKM matrix
given in Eq.~(7) is symmetric, a property that may not be fulfilled by 
present experimental data if we assume the unitarity of the CKM 
matrix. For this reason we will assume that the matrix
$\hat A$ has the form given in Eq.~(9) only at the scale of GU. The CKM
matrix at the scale of 1 GeV will then be obtained by using the method of the
RG and is not symmetric.

\section{Scheme for the generation of the CKM matrix}
Using the ideas outlined in the previous sections we will present here the
explicit construction of the CKM matrix. As we discussed the matrices of the
Yukawa couplings have a especially simple form after the rephasing of the
quark fields. The transformation of the rephasing that we will use is the
following 
\begin{equation}
\left( 
\begin{array}{c}
u \\ 
c \\ 
t
\end{array}
\right) _L\rightarrow \left( 
\begin{array}{ccc}
e^{-\frac \pi 3i} & 0 & 0 \\ 
0 & e^{\frac \pi 3i} & 0 \\ 
0 & 0 & 1
\end{array}
\right) \left( 
\begin{array}{c}
u \\ 
c \\ 
t
\end{array}
\right) _L,\ \left( 
\begin{array}{c}
d \\ 
s \\ 
b
\end{array}
\right) _L\rightarrow \left( 
\begin{array}{ccc}
e^{\frac \pi 3i} & 0 & 0 \\ 
0 & e^{-\frac \pi 3i} & 0 \\ 
0 & 0 & 1
\end{array}
\right) \left( 
\begin{array}{c}
d \\ 
s \\ 
b
\end{array}
\right) _L.  \nonumber
\end{equation}
Upon this transformation only the charged current and the Yukawa couplings
are changed. The matrix of the charged weak current becomes $D$ as given in 
Eq.~(8), and we choose the Yukawa couplings at the GU scale to be 
\begin{eqnarray}
\frac{v}{m_b\sqrt{2}}\,\,Y_d\sim \left( 
\begin{array}{ccc}
\alpha _{11}\lambda ^4 & \alpha _{12}\lambda ^3 & \frac{A\lambda ^3}3 \\ 
\alpha _{21}\lambda ^3 & \alpha _{22}\lambda ^2 & \frac{A\lambda ^2}{\sqrt{3}%
} \\ 
\frac{A\lambda ^3}3 & \frac{A\lambda ^2}{\sqrt{3}} & 1
\end{array}
\right) ,
\end{eqnarray}
where $v$ is the vacuum expectation value of the Higgs field. The 
corresponding structure of $Y_u$ is determined by Eq.~(11).
The parameters $A$ and $\lambda $ are the same as those of Wolfenstein's 
parameterization 
and only the leading powers of $\lambda $ are displayed. $\alpha_{ij}$ 
are numerical constants of ${\cal O}(1)$ which specific values are 
not essential for our discussion. Indeed, the structure of the CKM matrix 
will emerge only from the hierarchy of the third row and column of $Y_{u,d}$.

The matrix $\hat A$ that diagonalizes the matrices $Y_u$ and $Y_d$ has the
form 
\begin{equation}
\displaystyle \hat A=\left( 
\begin{array}{ccc}
\displaystyle n_1 & \displaystyle -\frac \lambda {{\sqrt{3}}}n_1 & %
\displaystyle 0 \\ 
\displaystyle\frac \lambda {{\sqrt{3}}}n_2 & \displaystyle n_2 & %
\displaystyle -\frac{A{\lambda }^2}{{\sqrt{3}}}n_2 \\ 
\displaystyle\frac{A{\lambda }^3}3n_3 & \displaystyle \frac{A{\lambda }^2}{{%
\sqrt{3}}}n_3 & \displaystyle \left( 1+\frac{{\lambda }^2}3\right) n_3
\end{array}
\right) 
\end{equation}
where $\frac 1{n_1}=\sqrt{1+\frac{|\lambda |^2}3},\quad \frac 
1{n_2}=\sqrt{1+%
\frac{|\lambda |^2}3+\frac{|A|^2|\lambda |^4}3},\quad $and $\quad 
n_3=n_1n_2$. The
matrix $\hat A$ given in Eq.~(14) is unitary and real and leads to the
symmetrical CKM matrix 
\begin{eqnarray}
V_{ckm}=\hat AD\hat A^T.
\end{eqnarray}
with entries given by
\[
\left| V_{11}\right| =n_1^2\left| e^{-\frac{2\pi }3i}\left( 1+\frac{\lambda
^2}3e^{\frac{4\pi }3i}\right) \right| \approx 1
\]
\[
\left| V_{12}\right| =n_1n_2\left| \frac \lambda {\sqrt{3}}e^{-\frac{2\pi }%
3i}\left( 1-e^{\frac{4\pi }3i}\right) \right| \approx \lambda 
\]
\[
\left| V_{13}\right| =n_1n_3\frac{A\lambda ^3}3\left| e^{-\frac{2\pi }%
3i}\left( 1-e^{\frac{4\pi }3i}\right) \right| \approx \frac{A\lambda ^3}{%
\sqrt{3}}
\]
\[
\left| V_{22}\right| =n_2^2\left| e^{\frac{2\pi }3i}+\frac{\lambda ^2}3e^{-%
\frac{2\pi }3i}+\frac{A^2\lambda ^4}3\right| \approx 1
\]
\[
\left| V_{23}\right| =n_2n_3\frac{A\lambda ^2}{\sqrt{3}}\left| e^{\frac{2\pi 
}3i}+\frac{\lambda ^2}3e^{-\frac{2\pi }3i}-\left( 1+\frac{{\lambda }^2}%
3\right) \right| \approx A\lambda ^2
\]
so it reproduces the Wolfenstein's hierarchy of the CKM matrix with fixed
values of $\rho $ and $\eta $. Moreover the 
matrix (15) already includes the CP violating phase and the ``plaquette''
\begin{equation}
J(GU)=\mbox{Im}\left( V_{11}V_{22}V_{12}^{*}V_{21}^{*}\right) 
=\frac{A^2\lambda ^6}{2\sqrt{3}}\approx (2 \sim 3)\cdot 10^{-5},
\end{equation}
has the right order of magnitude (see Table 1).

Now we perform the RG evolution of the matrices $Y_u$ and 
$Y_d$ from the GU energy to the scale of 1 GeV. The RG
equations become modified \cite{next} by the quark field rephasing
transformation given in Eq.~(12) and the evolution of the matrices $Y_u$ and
$Y_d$ are different.

 At the GU scale, the matrices $Y_u$ and $Y_d$ depend on the 
parameters $A$
and $\lambda $ that also define the common diagonalization matrix, Eq.~(14). 
Upon
the RG evolution, the matrices that diagonalize the Yukawa
couplings $Y_u$ and $Y_d$ evolve in such a way that the values of $A$ and $%
\lambda $ do not change for $Y_u$ and  they become complex for $Y_d$. At
the scale of 1 GeV $\tilde\lambda$ and $\tilde A$ are equal to 
\begin{eqnarray}
\tilde \lambda  &=&\lambda \left( 1\ GeV\right) =\frac{\left( 1+R
e^{-2i\pi /3}\right) }{\left( 1+Re^{2i\pi /3}\right) 
}\lambda \left( GU_{scale}\right) ,  \nonumber  \label{c52} \\
\tilde A &=&A\left( 1\ GeV\right) =\frac{a\left( 1+R
e^{2i\pi /3}\right) ^3}{\left( 1+Re^{-2i\pi
/3}\right) ^2}A\left( GU_{scale}\right).
\end{eqnarray}
Here $R$ and $a$ are the coefficients that depend on
the model that is used. Their form follows from the RG
equations and is given in Ref.~\cite{next}. The matrix that diagonalizes  
$Y_u$ at the scale of 1 GeV is thus equal to $\hat A$ given in Eq.~(14) and
the matrix that diagonalizes $Y_d$ is 
\begin{equation}
\displaystyle \hat A_d=\left( 
\begin{array}{ccc}
\displaystyle n_1 & \displaystyle -\frac{\tilde \lambda ^{*}}{{\sqrt{3}}}n_1
& \displaystyle 0 \\ 
\displaystyle\frac{\tilde \lambda }{{\sqrt{3}}}n_2 & \displaystyle n_2 & %
\displaystyle -\frac{\tilde A^{*}\left( {\tilde \lambda ^{*}}\right) ^2}{{%
\sqrt{3}}}n_2 \\ 
\displaystyle\frac{\tilde A{\tilde \lambda }^3}3n_3 & \displaystyle \frac{%
\tilde A{\tilde \lambda }^2}{{\sqrt{3}}}n_3 & \displaystyle \left( 1+\frac{%
\left| {\tilde \lambda }\right| ^2}3\right) n_3
\end{array}
\right).
\end{equation}
Using the matrices $\hat A$ and $\hat A_d$ the CKM matrix at the scale of 1
GeV becomes 
\begin{equation}
\hat V_{ckm}=\hat AD\hat A_d^{\dagger }.
\end{equation}
The matrix (19) is not symmetric and the pattern of non-symmetry of 
the CKM matrix is the same as in Wolfenstein's parameterization, Eq.~(6).

  In what follows, the parameters $A$ and $\lambda$ defined at the GU 
scale are adjusted so as to reproduce the CKM matrix at low energies. In 
order to illustrate their evolution to low energies, we have used the SM 
and its two Higgs (DHM) and minimal supersymmetric (MSSM) extensions. 
The experimental data on the CKM matrix are taken from 
Ref.~\cite{PDG} and the summary of our results is given in Table~1. 
\begin{table}[ht]
\begin{tabular}{|c|c|c|c|c|}
\hline
&Experiment&MSSM&SM&DHM\\ \hline
$\chi^2$&&11.47&5.60&10.34\\ \hline
$A\pm\Delta A$&&0.607$\pm$0.045&1.397$\pm$0.104 & 0.682$\pm$0.051
\\ \hline
$\lambda \pm \Delta \lambda $&&0.2433$\pm$0.0019 & 0.1984$\pm$0.0015&
0.2363$\pm$0.0018 \\ \hline
$|V_{ud}|$&0.9736$\pm$0.0010&0.97517&0.97514&0.97517\\ \hline
$|V_{us}|$&0.2205$\pm$0.0018&0.22138&0.22153&0.22141\\ \hline
$|V_{cd}|$&0.224$\pm$0.016&0.22141&0.22146&0.22142\\ \hline
$|V_{ub}|$& -- &0.00571&0.00467&0.00554\\ \hline
$|V_{td}|$& -- &0.00455&0.00699&0.00482\\ \hline
$\frac{|V_{ub}|}{|V_{cb}|}$&0.08$\pm$0.02&0.1391&0.1138&0.1352\\ \hline
$|V_{cb}|$&0.041$\pm$0.0030&0.0410&0.0410&0.0410\\ \hline
\end{tabular}
\caption{Comparison of the CKM matrix results for 
various models. Notice that the fitted values of $A$ and $\lambda$ are
at the GU scale while $|V_{ij}|$ correspond to their low energy values.}
\end{table}
From Table~1 we observe that the asymmetry of CKM matrix elements are of 
the expected order of magnitude (see Eq.~(6)). 
 The main difference in the results lies in the fact 
that $|V_{ub}|<|V_{td}|$ in the standard model, while $|V_{ub}|>|V_{td}|$ 
for the other two models. Since the values of $A$ and $\lambda$ are 
essentially fixed from $|V_{us}|$ and $|V_{cb}|$, the rather large value 
for $|V_{ud}|$ arises from unitarity of the CKM matrix. Although neither 
of the considered models 
can be really excluded, observe that the ratio $|V_{ub}/V_{cb}|$ is 
closest to its experimental value in the case of the 
SM and that a better fit\footnote{ 
There is a large contribution to $\chi^2$ that has its origin in the
fact that the values of $\left| V_{ub}\right| $ calculated from unitarity
($\left| V_{ub}\right| =0.059\pm 0.018$) and from $\left| V_{ub}\right| 
=\left|
V_{cb}\right| \frac{\left| V_{ub}\right| }{\left| V_{cb}\right| }=0.0033\pm
0.0009$ differ by 3 standard deviations. The data are thus incompatible at
this level with unitarity and any model that preserves the unitarity of CKM
matrix will show a corresponding discrepancy.},
reflected in the lowest $\chi^2$ value, is also obtained in this case.

From Eq.~(19) we can also compute the value of Jarlskog's parameter 
(Eq.~(16)) at low energies. The change in the value of $J$ is, as 
expected \cite{gilman}, negligibly small (at most 3 \% in the case of the 
SM) when evolving from GU to low energy scales.

\section{Discussion of the results and conclusions} 
We have presented a new kind of symmetry for the up and down
quark mass matrices that generates a CKM matrix in excellent
agreement with the experimental data. Our method is not based on
textures (or zeros) for the Yukawa couplings at GUT scales, but on the 
assumption that the symmetry of the CKM matrix is intimately related 
to a simple relation between Yukawa couplings of up and down quarks (see 
Eq.~(11)). Thus, we propose that at GUT scales $ Y_u=CY_d^2$ (which is valid 
for a special choice of quark fields phases) where $C$ can be estimated to be
\begin{equation}
C^{-1}=\frac{\sqrt{2}m_b^2}{vm_t}\approx5.4\cdot10^{-4}\approx\lambda^{5},
\label{eq25}
\end{equation}
Unfortunately we cannot say what kind of physics might be behind the relation
given in Eq.~(\ref{eq11}) but its simplicity and excellent predictions make
it a very attractive scenario for the generation of the CKM matrix.

It might seem that our method does not depend on the values
of the quark masses. This is true only to some extent. Our method only works
if there is a hierarchy for the quark masses and in such a case
the values of the quark masses are irrelevant. The form of the CKM matrix is 
entirely determined only by the third row and column of the $Y_u$ and 
$Y_d$ matrices, which suggests the essential role played by the 
heaviness of the third generation of quarks. We will
discuss the problem of the quark masses elsewhere \cite{next}. Let us only 
mention that we can also impose textures and, in such a way, reduce the
number of parameters still further.

One might ask if the method based on textures is equivalent
to ours. To examine this one has to check the relation (11) 
for the matrices $Y_u$ and $Y_d$ with given textures. We found that
Eq.~(11) is not fulfilled in any of the known schemes with textures.

The advantages of our method are the following
\begin{enumerate}
\item
It is based on a new symmetry between the $Y_u$ and $Y_d$
matrices which manifests itself that they are real
and diagonalizable by the same orthogonal matrix at
the GU scale.
\item
CP violation appears thanks to the phase factors
that are included in the charged weak current already
at the GU scale. These factors are not fitted and are
obtained from the condition $\hat{V}_{ckm}^3=1$. 
\item 
We reproduce with high accuracy the CKM matrix including 
the small asymmetry of the order $\lambda^6$. This last asymmetry
is the consequence of the RG
evolution from the GU scale to 1~GeV.
\item
Our method is very stable with respect to small changes in the values
of the initial data.
\end{enumerate}
For all these reasons this new symmetry of the $Y_u$
and $Y_d$ matrices may be a very important piece of information about
physics that lies at the basis of the standard model.

\noindent{\bf Acknowledgments}\\[2pt]
We are grateful to A.~Garc{\'\i}a, J.-M. G\'erard and J.~Pestieau for useful 
comments. S.R.J.W.\ gratefully acknowledges partial support by Comisi\'on
de Operaci\'on y Fomento de Actividades Acad\'emicas (Instituto
Polit\'ecnico Nacional). 


\begin{thebibliography}{99}
\bibitem{Fritzsch}  H.\ Fritzsch, Phys.\ Lett.\ {\bf B70}, 436 (1977); {\bf 
B73}, 317 (1978).

\bibitem{PDG} R.M.\ Barnett {\em et al.}, {\sl Particle Data Group}
Phys.\ Rev.\ {\bf D54}, Part I, (1996).

\bibitem{Dimop}  S.\ Dimopoulos, L.J.\ Hall, S.\ Raby, Phys.\ Rev.\ Lett.\ 
{\bf 68}, 1984 (1992); Phys.\ Rev.\ {\bf D45}, 4192 (1992); G.\ Anderson, S.\ 
Raby, S.\ Dimopoulos and L.J.\ Hall, Phys.\ Rev.\ {\bf D47}, 3702 (1993).

\bibitem{Ramond}  P.\ Ramond, R.G.\ Roberts and G.G.\ Ross, Nucl.\ Phys.\ {\bf 
B406}, 19 (1993).

\bibitem{Binetruy}  P.\ Binetruy and P.\ Ramond, Phys.\ Lett.\ {\bf B350}, 49 
(1995). 

\bibitem{Dudas}  E.\ Dudas, S.\ Pokorski and C.A.\ Savoy, Phys.\ Lett.\ {\bf 
B356}, 45 (1995).

\bibitem{Frampton}  P.H.\ Frampton and C.W.\ Kong, Phys.\ Rev.\ Lett.\ 
{\bf 75}, 78 (1995).

\bibitem{Pec-W}  R.D.\ Peccei and K.\ Wang, Phys.\ Rev.\ {\bf D53}, 2712 (1996).

\bibitem{Wang}  K.\ Wang, Phys.\ Rev.\ {\bf D54}, 5750 (1996).

\bibitem{Wolfen}  L.\ Wolfenstein, Phys.\ Rev.\ Lett.\ {\bf 51}, 1945 (1983).

\bibitem{Piotr}  P.\ Kielanowski, Phys.\ Rev.\ Lett.\ {\bf 63}, 2189 (1989). 

\bibitem{xing}  T.\ Kobayashi and Z.-Z.\ Xing, Mod.\ Phys.\ Lett.\ {\bf A11}, 
2531 (1996).

\bibitem{next}  H.\ Gonz\'alez, S.R.\ Ju\'arez W., P.\ Kielanowski and G.\
L\'opez Castro, to be published.

\bibitem{gilman} G.C.\ Athanasiu, S.\ Dimopoulos and F.J.\ Gilman, 
Phys.\ Rev.\ Lett.\ {\bf 57}, 1982 (1986).

\end{thebibliography}
\end{document}